\documentclass[showpacs,preprintnumbers,amsmath,amssymb]{revtex4}
\usepackage{graphicx}
\usepackage{dcolumn}
\usepackage{bm}

\newcommand{\ThreeJ}[6]{\left (\begin{array}{ccc}#1&#2&#3\\
#4&#5&#6\end{array} \right )}
\newcommand{\SixJ}[6]{\left \{\begin{array}{ccc}#1&#2&#3\\
#4&#5&#6\end{array} \right \}}

\begin{document}

\preprint{OU-TAP-268}
\preprint{YITP-06-06}

\title{Angular Trispectrum of CMB Temperature Anisotropy
from Primordial Non-Gaussianity with the Full Radiation Transfer Function}

\author{Noriyuki Kogo}
 \email{kogo@yukawa.kyoto-u.ac.jp}
\affiliation{Department of Earth and Space Science,
       Graduate School of Science, Osaka University,
       Toyonaka 560-0043, Japan}
\affiliation{Yukawa Institute for Theoretical Physics,
       Kyoto University, Kyoto 606-8502, Japan}
\author{Eiichiro Komatsu}
 \email{komatsu@astro.as.utexas.edu}
\affiliation{Department of Astronomy,
       The University of Texas at Austin,
       1 University Station, C1400, Austin, TX 78712}

\begin{abstract}
We calculate the cosmic microwave background (CMB) angular trispectrum,
spherical harmonic transform of the four-point correlation function,
from primordial non-Gaussianity in primordial curvature perturbations
characterized by a constant non-linear coupling parameter, $f_{\rm NL}$.
We fully take into account the effect of the radiation transfer function,
and thus provide the most accurate estimate of the signal-to-noise ratio
of the angular trispectrum of CMB temperature anisotropy.
We find that the predicted signal-to-noise ratio of the trispectrum
summed up to a given $l$ is approximately a power-law,
$(S/N)(<l)\sim 2.2\times 10^{-9}f^2_{\rm NL}l^2$,
up to the maximum multipole that we have reached in our numerical calculation,
$l=1200$, assuming that the error is dominated by cosmic variance.
Our results indicate that the signal-to-noise ratio of the temperature
trispectrum exceeds that of the bispectrum at the critical multipole,
$l_c \sim 1500~(50/|f_{\rm NL}|)$.
Therefore, the trispectrum of the Planck data is more sensitive
to primordial non-Gaussianity than the bispectrum for $|f_{\rm NL}|\gtrsim 50$.
We also report the predicted constraints on the amplitude of trispectrum,
which may be useful for other non-Gaussian models such as curvaton models.
\end{abstract}

\maketitle

Inflation has been the standard paradigm
for the origin of cosmological fluctuations.
While simple inflationary models based on a slowly-rolling scalar field
are unable to generate the detectable level of primordial non-Gaussianity,
a large class of models predict much stronger signals
(see \cite{BKMR04} for a review).
Non-Gaussianity of the primordial fluctuations thus
plays an important role in testing and constraining inflationary models.
As different statistical methods
are sensitive to different aspects of non-Gaussianity,
one should explore a variety of methods
in order to maximize our sensitivity to primordial non-Gaussianity.
Of which, the higher-order correlation functions such as
the three- and four-point correlation functions,
or their harmonic counterparts,
the bispectrum \cite{KS01,K02,KSW05,BCZ04,BZ04,B05,L05,C05}
and trispectrum \cite{KTHESIS,KBCFG01,H01,OH02},
have been actively investigated in the literature
as a powerful probe of primordial non-Gaussianity.

Compared with progress in theoretical calculations
of the angular bispectrum of CMB temperature \cite{KS01,L05}
and polarization \cite{BZ04} anisotropy,
that for the angular trispectrum has been a little bit behind.
The first calculation done by \cite{OH02} did not include
the full effect of radiation transfer function
caused by acoustic physics at the surface of last scatter.
In this paper, we calculate
the angular trispectrum of CMB temperature anisotropy,
fully taking into account the radiation transfer function.
Our estimate of the signal-to-noise ratio of the trispectrum should
thus improve accuracy of the previous estimate based on an approximate method.
We then compare the signal-to-noise ratios of primordial non-Gaussianity
from the trispectrum and bispectrum.

Let us briefly review the formalism of the CMB angular trispectrum,
following \cite{H01,OH02}.
We decompose temperature anisotropy on the sky, $\delta T/T$,
into spherical harmonic coefficients, $a_{lm}$, as
\begin{eqnarray}
\frac{\delta T}{T}(\hat{\bm{n}})
=\sum_{l m}a_{l m}Y_{l m}(\hat{\bm{n}}).
\label{TEMP}
\end{eqnarray}
Statistical isotropy of the universe requires
an $n$-point correlation function be rotationally invariant;
thus, for $n=4$ one obtains
\begin{eqnarray}
\langle a_{l_1 m_1}a_{l_2 m_2}a_{l_3 m_3}a_{l_4 m_4} \rangle
=\sum_{LM}(-1)^M
\ThreeJ{l_1}{l_2}{L}{m_1}{m_2}{-M}\ThreeJ{l_3}{l_4}{L}{m_3}{m_4}{M}
T^{l_1 l_2}_{l_3 l_4}(L),
\label{TRIDEF}
\end{eqnarray}
where $T^{l_1 l_2}_{l_3 l_4}(L)$ is the angular averaged trispectrum, 
$L$ is the length of a diagonal that forms triangles 
with $l_1$ and $l_2$ and with $l_3$ and $l_4$,
and the matrix is the Wigner 3-$j$ symbol,
which guarantees that two sides and the diagonal form a triangle,
$|l_1-l_2| \le L \le l_1+l_2$ and $|l_3-l_4| \le L \le l_3+l_4$.
Parity invariance also requires $l_1+l_2+L={\rm even}$, $l_3+l_4+L={\rm even}$,
$m_1+m_2-M=0$, and $m_3+m_4+M=0$.
These conditions determine the number of possible configurations.

The trispectrum generically consists of the connected, $T_c$,
and unconnected, $T_G$, part:
\begin{eqnarray}
T^{l_1 l_2}_{l_3 l_4}(L)
=T_c{}^{l_1 l_2}_{l_3 l_4}(L)+T_G{}^{l_1 l_2}_{l_3 l_4}(L).
\label{TRIDIV}
\end{eqnarray}
The former contains non-Gaussian signatures,
while the latter contains only the angular power spectrum,
$C_l \equiv \langle |a_{l m}|^2 \rangle$,
\begin{eqnarray}
T_G{}^{l_1 l_2}_{l_3 l_4}(L)
&=&(-1)^{l_1+l_3}\sqrt{(2l_1+1)(2l_3+1)}C_{l_1}C_{l_3}
   \delta_{l_1 l_2}\delta_{l_3 l_4}\delta_{L0} \nonumber \\
&&+(2L+1)C_{l_1}C_{l_2}
   \left[ (-1)^{l_1+l_2+L}
          \delta_{l_1 l_3}\delta_{l_2 l_4}
         +\delta_{l_1 l_4}\delta_{l_2 l_3} \right].
\label{TRIG}
\end{eqnarray}
Using permutation symmetry,
one may write the connected part of the trispectrum as
\begin{eqnarray}
T_c{}^{l_1 l_2}_{l_3 l_4}(L)
=P^{l_1 l_2}_{l_3 l_4}(L)
+(2L+1)\sum_{L'}
   \left[ (-1)^{l_2+l_3}\SixJ{l_1}{l_2}{L}{l_4}{l_3}{L'}
   P^{l_1 l_3}_{l_2 l_4}(L')
+(-1)^{L+L'}\SixJ{l_1}{l_2}{L}{l_3}{l_4}{L'}
   P^{l_1 l_4}_{l_3 l_2}(L') \right ],
\label{TRIC}
\end{eqnarray}
where
\begin{eqnarray}
P^{l_1 l_2}_{l_3 l_4}(L)
=t^{l_1 l_2}_{l_3 l_4}(L)
+(-1)^{2L+l_1+l_2+l_3+l_4}t^{l_2 l_1}_{l_4 l_3}(L)
+(-1)^{L+l_3+l_4}t^{l_1 l_2}_{l_4 l_3}(L)
+(-1)^{L+l_1+l_2}t^{l_2 l_1}_{l_3 l_4}(L).
\label{TRIP}
\end{eqnarray}
Here, the matrix is the Wigner 6-$j$ symbol,
and $t^{l_1 l_2}_{l_3 l_4}(L)$ is called the reduced trispectrum,
which contains all the physical information about non-Gaussian sources.

We parameterize primordial Bardeen's curvature perturbation, $\Phi$,
during the matter-dominated era in the usual form as \cite{KS01,OH02}
\begin{eqnarray}
\Phi(\bm{x})=\Phi_{\rm L}(\bm{x})
+f_{\rm NL}\left[ \Phi_{\rm L}^2(\bm{x})
-\langle \Phi_{\rm L}^2(\bm{x}) \rangle \right]
+f_2\Phi_{\rm L}^3(\bm{x}),
\label{PHI}
\end{eqnarray}
where $\Phi_{\rm L}$ is the linear Gaussian part
and $f_{\rm NL}$ and $f_2$ are the non-linear coupling parameters.
(Note that $\Phi$ is a perturbation in the $(i,i)$-component of the metric.)
The current observational constraint on $f_{\rm NL}$ is $-58<f_{\rm NL}<134$,
which is from the analysis
of the angular bispectrum of the WMAP data \cite{WMAPGAUSS1,WMAPGAUSS2}.
No constraint on $f_{\rm NL}$ from the angular trispectrum
is currently available.
It is therefore important to obtain an accurate estimate
of the signal-to-noise ratio of the angular trispectrum
of primordial non-Gaussianity expected from the WMAP
as well as Planck experiments \cite{PLANCK}.
While we take $f_{\rm NL}$ to be a constant independent of
coordinates or wavenumbers, in general $f_{\rm NL}$ depends on scales,
and different inflationary models predict different dependence of $f_{\rm NL}$
on wavenumbers \cite{BKMR04}.
The post-inflationary evolution of $\Phi$
due to second-order metric perturbations,
which must exist in any models of the standard cosmology,
also generates wavenumber-dependent $f_{\rm NL}$ \cite{L05}.
Nevertheless, a constant $f_{\rm NL}$ model is still useful
for estimating sensitivity of CMB experiments
to the amplitude of non-Gaussianity.
Measurements of non-Gaussian fluctuations are usually quite challenging
as non-Gaussianity is very small,
which makes detection of wavenumber-dependent features even more challenging.
Since currently there is no prediction for sensitivity of CMB experiments
to $f_{\rm NL}$ from the angular trispectrum of CMB temperature anisotropy
with the full radiation transfer function taken into account,
we shall adopt a constant $f_{\rm NL}$ to explore the signal-to-noise ratio
of the trispectrum as a function of the maximum multipoles
measured by observations.
In the future one may extend our approach
to include scale-dependent $f_{\rm NL}$,
following the method given in \cite{L05}, for instance.

The primordial fluctuations yield temperature anisotropy as
\begin{eqnarray}
a_{lm}=4\pi(-i)^l\int\frac{d^3\bm{k}}{(2\pi)^3}
\Phi(\bm{k})g_{Tl}(k)Y_{lm}^{*}(\hat{\bm{k}}),
\label{ALM}
\end{eqnarray}
where $g_{Tl}(k)$ is the radiation transfer function
of adiabatic fluctuations, which can be calculated numerically
by the CMBFAST code \cite{SZ96}. 
Using this relation,
we obtain the formula of the reduced angular trispectrum as
\begin{eqnarray}
t^{l_1 l_2}_{l_3 l_4}(L)
&=&\int r_1^2dr_1 r_2^2dr_2~F_L(r_1,r_2)
   \alpha_{l_1}(r_1)\beta_{l_2}(r_1)\alpha_{l_3}(r_2)\beta_{l_4}(r_2)
   h_{l_1 L l_2}h_{l_3 L l_4} \nonumber\\
&&+\int r^2 dr~\beta_{l_2}(r)\beta_{l_4}(r)
   \left[ \mu_{l_1}(r)\beta_{l_3}(r)+\beta_{l_1}(r)\mu_{l_3}(r) \right]
   h_{l_1 L l_2}h_{l_3 L l_4},
\label{TRIPRIM}
\end{eqnarray}
where
\begin{eqnarray}
F_L(r_1,r_2)&\equiv&\frac{2}{\pi}\int k^2dk~ P_\Phi(k)j_L(kr_1)j_L(kr_2),
\label{FL} \\
 \alpha_l(r)&\equiv&\frac{2}{\pi}\int k^2dk~ (2f_{\rm NL})g_{Tl}(k)j_l(kr),
\label{AL} \\
  \beta_l(r)&\equiv&\frac{2}{\pi}\int k^2dk~ P_\Phi(k)g_{Tl}(k)j_l(kr),
\label{BL} \\
    \mu_l(r)&\equiv&\frac{2}{\pi}\int k^2dk~ f_2g_{Tl}(k)j_l(kr),
\label{UL}
\end{eqnarray}
and
\begin{eqnarray}
h_{l_1 L l_2}
\equiv\sqrt{\frac{(2l_1+1)(2l_2+1)(2L+1)}{4\pi}}\ThreeJ{l_1}{l_2}{L}{0}{0}{0}.
\label{HLLL}
\end{eqnarray}
We have defined the power spectrum of the primordial curvature perturbation as
\begin{eqnarray}
\langle \Phi_{\rm L}(\bm{k})\Phi_{\rm L}^*(\bm{k}') \rangle
=(2\pi)^3\delta^{(3)}(\bm{k}-\bm{k}')P_\Phi(k),
\label{PK}
\end{eqnarray}
where $P_\Phi\propto k^{n-4}$ and $n=1$ for a scale-invariant spectrum.

The signal-to-noise ratio of the temperature trispectrum is given by
\begin{eqnarray}
\left(\frac{S}{N}\right)_{\rm tri}^2
=\sum_{l_1>l_2>l_3>l_4}\sum_L
\frac{[T_c{}^{l_1 l_2}_{l_3 l_4}(L)]^2}
{(2L+1)C_{l_1}C_{l_2}C_{l_3}C_{l_4}}.
\label{SNTRI}
\end{eqnarray}
Note that we have assumed that the error is solely due to the cosmic variance.
When detector noise is included,
$C_l$ in the denominator should be replaced by $C_l+N_l/W_l$,
where $N_l$ is the noisebias and $W_l$ is the window function.
We set the normalization of $P_\Phi(k)$
by the height of the first acoustic peak of the WMAP data;
$l(l+1)C_l/2\pi=(74.7{\rm \mu K})^2$ at $l=220$,
and adopt the cosmological parameters of the concordance $\Lambda$CDM model
with a scale-invariant primordial spectrum for $\Phi$;
$h=0.72$, $\Omega_b=0.047$, $\Omega_\Lambda=0.71$, $\Omega_m=0.29$,
and $\tau=0.17$ \cite{SPERGEL}.
While we shall evaluate $(S/N)^2$ using the trispectrum
that uses the radiation transfer function [equation~(\ref{TRIPRIM})] below,
let us first estimate an order-of-magnitude of $(S/N)^2$
and its dependence on the maximum multipole, $l_{\rm max}$,
using the Sachs--Wolfe approximation valid at low multipoles ($l \ll 100$),
$g_{Tl}(k) \approx -j_l(kr_*)/3$,
where $r_*$ is the comoving distance to the surface of last scatter.
This form of $g_{Tl}(k)$ then gives
\begin{eqnarray}
t^{l_1 l_2}_{l_3 l_4}(L)
\approx 9C_{l_2}^{\rm SW}C_{l_4}^{\rm SW}\left[ 4f_{\rm NL}^2C_L^{\rm SW}
       +f_2\left( C_{l_1}^{\rm SW}+C_{l_3}^{\rm SW} \right) \right]
        h_{l_1 L l_2}h_{l_3 L l_4},
\label{TRISW}
\end{eqnarray}
where
\begin{eqnarray}
C_l^{\rm SW}=\frac{2}{9\pi}\int k^2dk P_\Phi(k)j_l^2(kr_*)
=\frac{A}{l(l+1)},
\label{CLSW}
\end{eqnarray}
is the angular power spectrum in the Sachs--Wolfe approximation
for a scale-invariant primordial spectrum of $\Phi$,
and $A \approx 6\times10^{-10}$.
For simplicity we take into account
only the first terms in equations~(\ref{TRIC}) and (\ref{TRIP}).
Moreover, let us consider only $L=1$ modes
because the collapsed configurations, which correspond to the low-$L$ modes, 
are the dominant modes for $f_{\rm NL}$ model [equation (\ref{PHI})].
This property also holds for the bispectrum \cite{BCZ04,C05}.
For $f_2=0$ we have
\begin{eqnarray}
t^{l_1 l_2}_{l_3 l_4}(L=1)
\approx 36f_{\rm NL}^2C_{L=1}^{\rm SW}C_{l_2}^{\rm SW}C_{l_4}^{\rm SW}
        h_{l_1,1,l_2}h_{l_3,1,l_4}.
\label{TRISWO}
\end{eqnarray}
Since $|l_1-1| \le l_2 \le l_1+1$ from the triangle condition and $l_1>l_2$,
$h_{l_1,1,l_2}$ is non-zero only for $l_1-1=l_2$. Hence
\begin{eqnarray}
h_{l_1,1,l_2}^2
&=&\frac{3}{4\pi}(2l_1+1)(2l_2+1)\ThreeJ{l_1}{l_2}{1}{0}{0}{0}^2 \nonumber \\
&=&\frac{3}{4\pi}(2l_1+1)(2l_1-1)\times\frac{l_1}{(2l_1-1)(2l_1+1)}
   \delta_{l_1-1,l_2} \nonumber \\
&=&\frac{3l_1}{4\pi}\delta_{l_1-1,l_2},
\label{HLLLO}
\end{eqnarray}
which gives
\begin{eqnarray}
\left(\frac{S}{N}\right)_{\rm tri}^2
&\approx& \sum_{l_1>l_2>l_3>l_4}\frac{[t^{l_1 l_2}_{l_3 l_4}(L=1)]^2}
       {3C_{l_1}^{\rm SW}C_{l_2}^{\rm SW}C_{l_3}^{\rm SW}C_{l_4}^{\rm SW}}
       \nonumber \\
&\approx& \left( \frac{27}{\pi} \right)^2 \frac{1}{4!} \sum_{l_1l_2l_3l_4}
       \frac{f_{\rm NL}^4C_{l_2}^{\rm SW}C_{l_4}^{\rm SW}(C_{L=1}^{\rm SW})^2
       l_1 l_3 \delta_{l_1-1,l_2}\delta_{l_3-1,l_4}}
       {3C_{l_1}^{\rm SW}C_{l_3}^{\rm SW}}
       \nonumber \\
&\approx& A^2f_{\rm NL}^4l_{\rm max}^4,
\label{SNTRIO}
\end{eqnarray}
where $l_{\rm max}$ is the maximum multipole.
This result indicates that $(S/N)^2$ of the trispectrum
strongly depends on both $f_{\rm NL}$ and $l_{\rm max}$,
$(S/N)^2_{\rm tri}\propto (f_{\rm NL}l_{\rm max})^4$,
which is stronger than that of the bispectrum,
$(S/N)^2_{\rm bi}\propto (f_{\rm NL}l_{\rm max})^2$ \cite{KS01}.
We have confirmed that $l_{\rm max}^4$ dependence still holds
when all the terms (still in the Sachs--Wolfe approximation) are included.
The constant of proportionality is however about 20 times larger
than equation~(\ref{SNTRIO}).
This is because there are other 11 terms due to the permutation symmetry
as seen in equations~(\ref{TRIC}) and (\ref{TRIP}),
and summing over $L>1$ modes further increases $(S/N)^2$ by a factor of two.
We shall show below that $l_{\rm max}^4$ dependence also holds very well
for the full calculations.

\begin{figure}
\begin{center}
\includegraphics[scale=0.5]{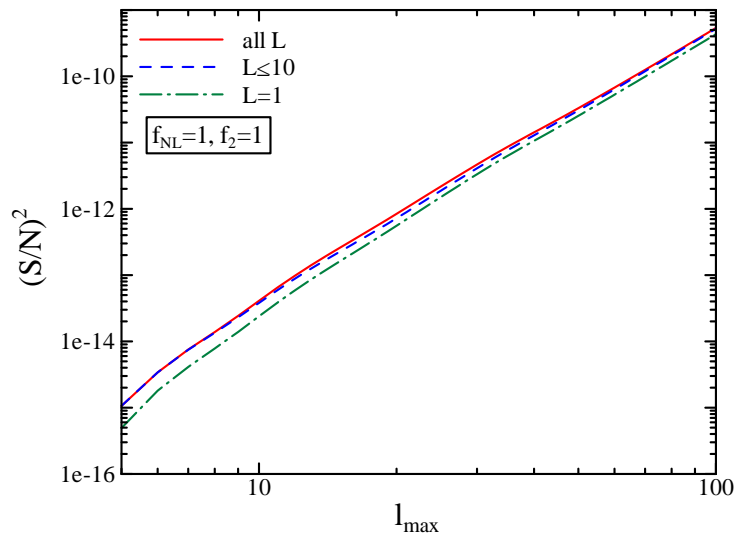}
\caption{Signal-to-noise ratio squared, $(S/N)^2$,
of the angular trispectrum summed over all $L$, $L \le 10$,
and only $L=1$ modes, respectively,
as a function of the maximum multipole, $l_{\rm max}$.
Here, $L$ denotes the multipole of the diagonal of a trispectrum configuration.
We assume $f_{\rm NL}=1$ and $f_2=1$
and use the full radiation transfer function.
This figure shows that the summation over $L$ needs to
be done only up to $L\simeq 10$.
\label{TRILIM}}
\end{center}
\end{figure}

\begin{figure}
\begin{center}
\includegraphics[scale=0.5]{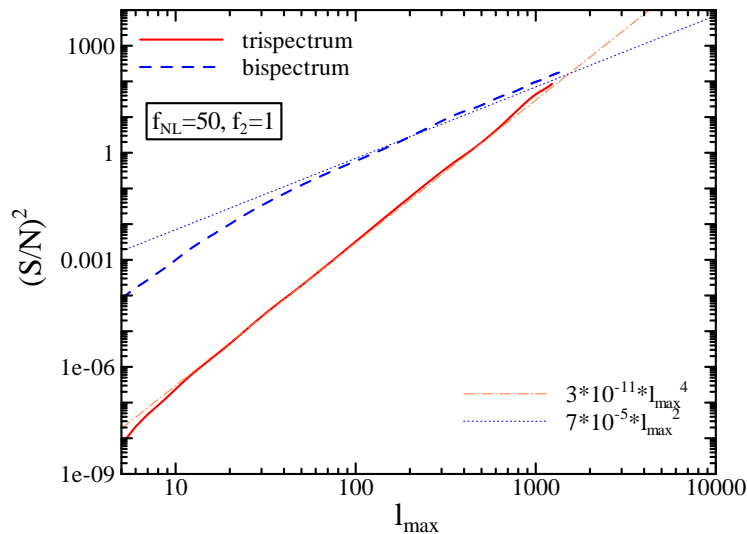}
\caption{Predicted signal-to-noise ratio squared, $(S/N)^2$,
of the angular trispectrum and bispectrum
with the full radiation transfer function for $f_{\rm NL}=50$ and $f_2=1$.
Power-law fits are also shown.
Note that the power-law fits break down at $l_{\rm max}\gtrsim 3000$,
where the gravitational lensing effects become important:
$(S/N)^2$ ceases to grow at $l_{\rm max}\sim 3000$ \cite{KS01,BZ04}.
\label{TRIBI}}
\end{center}
\end{figure}

As the trispectrum calculation involves the summation over five multipoles,
computation takes too long to let us go beyond $l_{\rm max}\sim 100$; however,
we find that almost all the contribution actually comes from $L \lesssim 10$
as shown in Fig.~\ref{TRILIM}
where we use the full radiation transfer function.
It is therefore sufficient to perform the summation over the diagonal, $L$,
only up to 10, which makes the computational time scale
as $l_{\rm max}^4$ instead of $l_{\rm max}^5$,
giving a huge saving in computational time.
Figure~\ref{TRIBI} shows the predicted signal-to-noise ratio squared,
$(S/N)^2$, for the trispectrum (summed over $L \le 10$)
as well as the bispectrum as a function of the maximum multipole,
$l_{\rm max}$, with the full radiation transfer function.
We have assumed a fiducial value of $f_{\rm NL}=50$ and $f_2=1$
in which case the effect of $f_2$ is negligible.
Our result suggests that the expected signal-to-noise ratio of
the primordial trispectrum is approximately given by
\begin{equation}
\left(\frac{S}{N}\right)^2_{\rm tri}
\simeq 5\times 10^{-18}~f_{\rm NL}^4l_{\rm max}^4,
\label{SNTRIN}
\end{equation}
or $(S/N)_{\rm tri}\simeq 2.2\times 10^{-9}~(f_{\rm NL}l_{\rm max})^2$.
The minimum detectable $f_{\rm NL}$ by the trispectrum at the 1-$\sigma$ level
is therefore $f_{\rm NL}\simeq 85$, 42, 21, 14, and 11
for $l_{\rm max}=250$, 500, 1000, 1500, and 2000.
This may be compared with the minimum detectable $f_{\rm NL}$
by the bispectrum, $f_{\rm NL}\simeq 23$, 12, 6, 4, and 3
for $l_{\rm max}=250$, 500, 1000, 1500, and 2000.
(Note again that we assume that the CMB bispectrum and trispectrum
are cosmic-variance limited up to $l_{\rm max}$.)
Here, we have used a power-law fit to
$(S/N)_{\rm bi}\simeq 1.7\times 10^{-4}~|f_{\rm NL}|l_{\rm max}$.
The power-law fits must break down at $l_{\rm max}\gtrsim 3000$,
where $(S/N)^2$ does not grow due to the gravitational lensing effects
increasing the cosmic variance \cite{KS01,BZ04}.

These results might seem to imply that the temperature trispectrum
is always less sensitive to $f_{\rm NL}$ than the bispectrum; however,
as $(S/N)_{\rm tri}$ depends on $f_{\rm NL}$ more strongly
than $(S/N)_{\rm bi}$, the ratio also depends on $f_{\rm NL}$:
\begin{equation}
\frac{(S/N)_{\rm tri}}{(S/N)_{\rm bi}}
\simeq \frac{|f_{\rm NL}|}{52}\frac{l_{\rm max}}{1500}.
\label{SNTRIR}
\end{equation}
The trispectrum actually becomes more sensitive
to primordial non-Gaussianity than the bispectrum for Planck-like experiments
probing $l_{\rm max}\gtrsim 1500$ and $|f_{\rm NL}|\gtrsim 50$.
Incidentally, the trispectrum cannot measure the sign of $f_{\rm NL}$,
as it depends on  $f_{\rm NL}^2$.
While we have considered the trispectrum of temperature anisotropy only,
the signal-to-noise ratio should increase when polarization is included
in the analysis.
The number of possible combinations
of the temperature and $E$-polarization field for the trispectrum is 5,
while that for the bispectrum is 4;
thus, one expects similar improvements in the signal-to-noise ratio
for the bispectrum and trispectrum.
The calculation including the temperature and polarization bispectra
has shown that the signal-to-noise ratio increases
roughly by a factor of 2 \cite{BZ04}.
We expect a similar level of improvement for the trispectrum as well.

Equation~(\ref{SNTRIR}) holds only for the particular non-Gaussian model
given by equation~(\ref{PHI});
however, the amplitude of the trispectrum may be related to
that of the bispectrum in a very different way for other models.
For example, the curvature perturbation may be given by
\begin{eqnarray}
\Phi(\bm{x})=\Phi_{\rm L}(\bm{x})
+f_\eta\left[ \eta^2(\bm{x})-\langle \eta^2(\bm{x}) \rangle \right],
\label{ETA}
\end{eqnarray}
instead of equation~(\ref{PHI}), where $\eta$ is another fluctuating field
that is totally uncorrelated with $\Phi_{\rm L}$.
This form of non-Gaussianity may arise from
a particular configuration of curvaton models \cite{LM97,LUW03}.
(The form of the CMB angular bispectrum arising from this model
has been derived in Appendix~C of \cite{KTHESIS}.)
In this model the ratio of the amplitude of trispectrum to bispectrum
is very different from that for equation~(\ref{PHI}).
Boubekeur and Lyth \cite{BL06} proposed to use the following parameterization
for the trispectrum amplitude of primordial curvature perturbation
in the comoving gauge, $\zeta$,
which is related to Bardeen's curvature perturbation during the matter era as
$\zeta=(5/3)\Phi$:
\begin{equation}
\langle\zeta(\bm{k}_1)\zeta(\bm{k}_2)\zeta(\bm{k}_3)\zeta(\bm{k}_4)\rangle
=(2\pi)^3\delta^{(3)}(\bm{k}_1+\bm{k}_2+\bm{k}_3+\bm{k}_4)\tau_{\rm NL}
 \left[P_\zeta(k_1)P_\zeta(k_2)P_\zeta(|\bm{k}_1+\bm{k}_4|)
 +(11~{\rm distinct~permutations})\right].
\label{TRIZETA}
\end{equation}
Equation~(\ref{PHI}) with $f_2=0$ gives $\tau_{\rm NL}=(6f_{\rm NL}/5)^2$,
while equation~(\ref{ETA}) gives different predictions \cite{BL06}.
Our results therefore give the minimum detectable $\tau_{\rm NL}$
at the 1-$\sigma$ level as
$\tau_{\rm NL}\simeq 10400$, 2500, 640, 280, and 170
for $l_{\rm max}=250$, 500, 1000, 1500, and 2000.
These results should be useful for estimating sensitivity of CMB experiments
to a variety of non-Gaussian models giving rise to non-zero trispectrum.

Our predictions show that the angular trispectrum measured from WMAP
is unable to detect primordial non-Gaussianity parameterized
by $f_{\rm NL}$ as in equation~(\ref{PHI}),
given the current constraint on $f_{\rm NL}$ from the bispectrum.
However, the trispectrum is at least as powerful as the bispectrum
for detecting primordial non-Gaussianity in the Planck data
if $|f_{\rm NL}|\sim 50$, which is still allowed by the WMAP data.
For other models, no detection of significant trispectrum
on the WMAP ($l_{\rm max}\sim 250$)
and Planck data ($l_{\rm max}\sim 1500$) would imply
$\tau_{\rm NL}\lesssim 2\times 10^4$ and 560 at the 2-$\sigma$ level,
respectively.

\begin{acknowledgments}
We would like to thank organizers of the Yukawa International Seminar 2005
(YKIS2005) for giving us an opportunity to start this project.
We would like to thank David H. Lyth for useful comments on
early versions of the paper.
N.K. would like to thank Misao Sasaki for helpful advice.
E.K. acknowledges support from the Alfred P. Sloan Foundation.
\end{acknowledgments}


\begin{thebibliography}{}

\bibitem{BKMR04}
  N. Bartolo, E. Komatsu, S. Matarrese and A. Riotto,
  Phys. Rept. {\bf 402}, 103 (2004)
\bibitem{KS01}
  E. Komatsu and D. N. Spergel,
  Phys. Rev. D {\bf 63}, 063002 (2001)
\bibitem{K02}
  E. Komatsu, B. D. Wandelt, D. N. Spergel, A. J. Banday and K. M. G\'orski,
  Astrophys. J. {\bf 566}, 19 (2002)
\bibitem{KSW05}
  E. Komatsu, D. N. Spergel and B. D. Wandelt,
  Astrophys. J. {\bf 634}, 14 (2005)
\bibitem{BCZ04}
  D. Babich, P. Creminelli and M. Zaldarriaga,
  JCAP {\bf 0408}, 009 (2004)
\bibitem{BZ04}
  D. Babich and M. Zaldarriaga,
  Phys. Rev. D {\bf 70}, 083005 (2004)
\bibitem{B05}
  D. Babich,
  Phys. Rev. D {\bf 72}, 043003 (2005)
\bibitem{L05}
  M. Liguori, F. K. Hansen, E. Komatsu, S. Matarrese and A. Riotto,
  Phys. Rev. D, in press (astro-ph/0509098)
\bibitem{C05}
  P. Cabella, F. K. Hansen, M. Liguori, D. Marinucchi, S. Matarrese,
  L. Moscardini and N. Vittorio,
  preprint (astro-ph/0512112)
\bibitem{KTHESIS}
  E. Komatsu,
  PhD Thesis, Tohoku University (2001) (astro-ph/0206039)
\bibitem{KBCFG01}
  M. Kunz, A. J. Banday, P. G. Castro, P. G. Ferreira and K. M. Gorski,
  Astrophys. J. {\bf 563}, L99 (2001)
\bibitem{H01}
  W. Hu,
  Phys. Rev. D {\bf 64}, 083005 (2001)
\bibitem{OH02}
  T. Okamoto and W. Hu,
  Phys. Rev. D {\bf 66}, 063008 (2002)
\bibitem{WMAPGAUSS1}
  E. Komatsu {\it et al.},
  Astrophys. J. Suppl. {\bf 148}, 119 (2003)
\bibitem{WMAPGAUSS2}
  P. Creminelli, A. Nicolis, L. Senatore, M. Tegmark and M. Zaldarriaga,
  preprint (astro-ph/0509029)
\bibitem{PLANCK}
  Planck website; http://www.rssd.esa.int/index.php?project=PLANCK
\bibitem{SZ96}
  U. Seljak and M. Zaldarriaga,
  Astrophys. J. {\bf 469}, 437 (1996)
\bibitem{SPERGEL}
  D. N. Spergel {\it et al.},
  Astrophys. J. Suppl. {\bf 148}, 175 (2003)
\bibitem{LM97}
  A. Linde and V. Mukhanov,
  Phys. Rev. D {\bf 56}, R535 (1997)
\bibitem{LUW03}
  D. H. Lyth, C. Ungarelli and D. Wands,
  Phys. Rev. D {\bf 67}, 023503 (2003)
\bibitem{BL06}
  L. Boubekeur and D. H. Lyth,
  preprint (astro-ph/0504046)
\end{thebibliography}
\end{document}